\pdfoutput=1
\documentclass[aps,onecolumn,rmp]{revtex4}
\usepackage{amsmath,amssymb,graphicx}

\usepackage{fancyhdr,bm}
\usepackage{color}
\usepackage[colorlinks=true, letterpaper=true, pdfstartview=FitV, linkcolor=blue, citecolor=blue, urlcolor=blue]{hyperref}
\setcitestyle{square,comma,numbers}
\begin{document}

\title{Spontaneous Chiral Symmetry Breaking in Bilayer Graphene}

\author{Fan Zhang}\email{zhang@utdallas.edu}
\affiliation{Department of Physics, University of Texas at Dallas, Richardson, Texas 75080, USA}

\begin{abstract}
Bilayer graphene and its thicker cousins with Rhombohedral stacking have attracted considerable attention
because of their susceptibility to a variety of broken chiral symmetry states.
Due to large density-of-states and quantized Berry phases near their gapless band touching points,
each spin-valley flavor spontaneously transfers charge between layers to yield opening of energy gaps in quasiparticle spectra and spreading of momentum-space Berry curvatures.
In this article we review the development of theories that predicted such chiral symmetry breaking and classified the possible topological many-body ground states, and the observations in recent experiments that are in reasonable agreement with these theories.
\end{abstract}
\maketitle
\tableofcontents

\newpage 

\section{Introduction to Chiral Graphene}
Success in exfoliating monolayer and few-layer graphene sheets from bulk graphite,
combined with progress in their epitaxial growth, has opened up a rich new topic in two-dimensional electron systems (2DES)~\cite{RMP-09}.
Graphene 2DES are remarkable for several different reasons.
The fact that they are truly two dimensional on an atomic length scale elevates 2DES physics
from the low-temperature world to the room-temperature world.
Furthermore, they are accurately described by very simple models over very wide energy ranges
and yet have electronic properties that can be qualitatively altered simply by stacking them in different arrangements,
and by adjusting external gate voltages or magnetic fields.
Lastly but not the least, it is relatively easy to access graphene samples and to purify them,
which practically promotes the experimental examinations of fascinating theories on graphene 2DES.

The basic building block of all graphene 2DES is the isolated monolayer,
which is described by a massless Dirac ${\bm k}\cdot{\bm p}$ Hamiltonian over a wide energy range~\cite{RMP-09}.
Special for graphene, the Dirac model is massless,
with two Weyl points~\cite{Weyl} of opposite chiralities located at valley $K$ and $K'$, i.e., the two inequivalent Brillouin zone corners.
The massless Dirac model has linear dispersions and chiral quasiparticles, and in the graphene case the chirality
refers to the alignment between the direction of ${\bm k}\cdot{\bm p}$ momentum and the direction of pseudospin associated with the A/B sublattice degree-of-freedom of graphene's honeycomb lattice.
Intriguingly, the two Weyl points at valley $K$ and $K'$ are protected by the translational symmetry, a chiral symmetry, and a Berry phase $\pm\pi$.
We will elaborate more on these two features below.

\begin{figure}[t!]
\centering{ \scalebox{0.45} {\includegraphics*{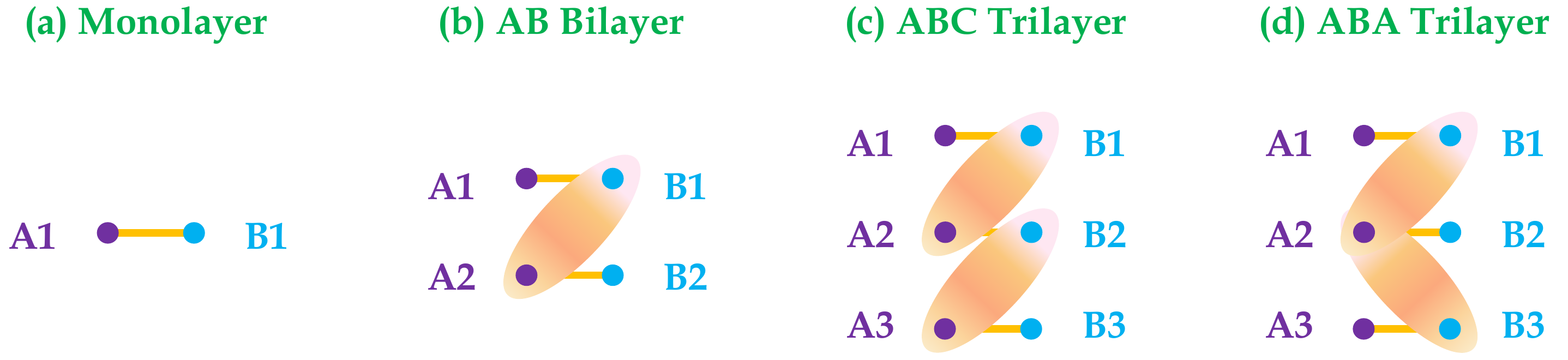}}}\caption{\label{stacking} {
Schematic electronic unit-cell structures of few-layer graphene. (a)-(c) Rhombohedral (ABC) stacked graphene layers.
The sublattices coupled by the strongest interlayer bonds characterized by the vertical $\gamma_1$ hopping parameter
are indicated by shading and have little weight in the low-energy effective states.
Both the monolayer and the bilayer can be viewed as either Rhombohedral (ABC) or Bernal (AB) stacked graphene.
(d) Bernal (ABA) stacked trilayer graphene for comparison. Figures adapted from ref.~\onlinecite{Zhang-12-QHF}.
The corresponding real-space stacking structures can be found in ref.~\onlinecite{RMP-09}.}}
\end{figure}

When $N$ honeycomb graphene layers are stacked, electronic properties are strongly modified in a way
that is controlled by the specific stacking arrangement~\cite{RMP-09}.
It turns out that among all the stacking possibilities, only the Rhombohedral (ABC or chiral) arrangement~\cite{Min-08-N,Koshino-09,Zhang-10-ABC}
inherits and extends the most interesting features of monolayer graphene, as we now explain~\cite{Zhang-10-ABC}.
(i) There are two low energy sublattice sites, as the other sublattice site energies are repelled from the Fermi level by the interlayer hopping $\gamma_1$ and thus irrelevant at low energies, as shown in Fig.~\ref{stacking}. This suggests that a two-band model provides a useful tool to describe the long-wavelength physics. (ii) The low-energy sublattice sites are localized in the outermost layers, at $A_1$ and $B_N$, and can be separated energetically by an electric field perpendicular to the film. (iii) Hopping between low-energy sites via high-energy states is an $N$-step process which leads to $\pm p^N$ dispersions in conduction and valence bands, and sublattice pseudospin chirality $N$ (or Berry phase $N\pi$).
(iv) The low-energy bands are increasingly flat for larger $N$, at least when weak remote hopping processes are neglected,
and the opportunity for interesting interaction and disorder physics is therefore stronger. Consequently, in the simplified chiral model, the
density-of-states $\nu(E)\sim E^{(2-N)/N}$ diverges as $E$ approaches zero for $N>2$ whereas it remains finite for $N=2$ and vanishes for $N=1$.
(These properties also have some relevance to more general stacking arrangements since the low-energy Hamiltonian of a multilayer with any type
of stacking can always be chiral-decomposed~\cite{Min-08-N} to a direct sum of ABC-stacked layers.
Monolayer and bilayer graphene can be viewed as ABC-stacked few-layers with $N=1$ and $N=2$, respectively.)

We refer the family of ABC-stacked $N$-layer graphene collectively as the chiral 2DES.
It follows the properties (i)-(iv) that the electronic properties of $N$-layer chiral 2DES can be well described by
$\bf{k} \cdot \bf{p}$ band Hamiltonians~\cite{Koshino-09,Zhang-10-ABC}
\begin{eqnarray}
\label{H0}
H_{N} =\frac{(v_0 p)^N}{(-\gamma_1)^{N-1}}\big[\cos(N \phi_{\bm p})\sigma_{x}+\sin(N \phi_{\bm p})
\sigma_{y}\big] \,.
\end{eqnarray}
We have used the notation $\cos\phi_{\bm p}=\tau_z p_{x}/p$ and $\sin\phi_{\bm p}=p_{y}/p$
where $\tau_{z}=\pm 1$ labels $K$ and $K'$ valleys.
The Pauli matrices ${\bm {\sigma}}$ act on a pseudospin degree-of-freedom, i.e., the two low-energy sublattices $A_1$ and $B_N$.
We choose the positive and negative eigenstates of $\sigma_z$ to denote $B_N$ (bottom layer) and $A_1$ (top layer), respectively.
The Pauli matrices $\bm s$ will be reserved to denote the electron spin.
$v_0\sim 10^6$~m/s is the Fermi velocity in graphene, and $\gamma_1\sim 0.4$ eV is the nearest neighbor interlayer hopping energy.
Neutral chiral 2DES with $N>1$ has been proved to be fertile ground for new many-body physics~\cite{Zhang-10-ABC,Zhang-12-QHF,Zhang-11-SQH}.
Because of the large density-of-states and the $N\pi$ Berry phases
near low energy band-contact points, such 2DES at zero external fields are susceptible to chiral symmetry breaking, leading to a family of gapped spontaneous quantum Hall states distinguished by valley and spin dependent quantized Hall conductivities~\cite{Zhang-11-SQH}.
In these states, each spin-valley flavor spontaneously transfers charge between layers~\cite{Zhang-11-SQH,Min-08-MF,Zhang-10-RG,Levitov-10-MF}.
Particularly in high mobility suspended bilayers~\cite{Bao-12}, reproducible experimental observations~\cite{Yacoby-10-1,Yacoby-10-2,Schonenberger-12,Lau-12,Wees-12,Schonenberger-13,Lau-14,Others} are in reasonable agreement with original theoretical predictions~\cite{Zhang-11-SQH,Min-08-MF,Zhang-10-RG,Levitov-10-MF}, both of which will be reviewed in this article.

\section{Semimetals with Protected Fermi Points}
Notably, the Fermi surface consists of two band touching points at $K$ and $K'$ for charge neutral chiral 2DES,
which are indeed protected.
As implied by Eq.~(\ref{H0}), the layer pseudospin rotates $N$ times faster than the momentum orientation angle.
This amounts to acquiring a Berry phase $N\pi$ when a quasiparticle circles one of the band-contact points once~\cite{Koshino-09,Zhang-10-ABC}.
The Berry phases are opposite for electron and hole bands, and for $K$ and $K'$ valleys.
The quantization of Berry phase, instead of being accidental, is directly dictated by the following chiral (sublattice) symmetry~\cite{Zhang-14-Chiral,Chiral-Note}
\begin{eqnarray}
\{H_{N}, \sigma_z\}=0\,.
\end{eqnarray}
This chiral symmetry requires that at any momentum $\bm p$ a state with energy $E$ must have a partner state with energy $-E$.
The gapless band-contact nature of the spectra of Hamiltonians~(\ref{H0}) is protected by the chiral symmetry,
since any loop enclosing one band touching point has a nontrivial Berry phase $\pm N\pi$ (a topological winding number $N$) and is thus not contractible.
One can always redefine the zero energy at each $\bm p$ to respect the chiral symmetry
if a $h({\bm p})\sigma_0$ term is introduced to Eq.~(\ref{H0}).
In this case, at each $\bm p$ the eigenstates do not alter and the Berry connection remains the same.
The chiral symmetry is also robust to any perturbation proportional to $\sigma_{x,y}$.
A notable example is the trigonal warping effect~\cite{Koshino-09,Zhang-10-ABC,McCann-06,Supp-12}, in which the additional $\sigma_{x,y}$ terms, instead of gapping a spectrum, only split a band-contact point with Berry phase $N\pi$ into Weyl points with Berry phase $\pm\pi$ each and $N\pi$ in total~\cite{Supp-12}.

However, the chiral symmetry is broken and the energy spectrum acquires a gap ($2m$ at $\bm p=0$) in the presence of a $m\sigma_z$ term~\cite{Supp-12}.
In this case, at each $\bm p$ the eigenstates becomes pseudospin polarized and the two monopoles at $\bm p=0$ spread out near the two valley centers, producing substantial momentum-space Berry curvatures.
Of course, even in the presence of the chiral symmetry, the band-contact points can be gapped out, if gauge symmetry or translational symmetry are broken. As two examples, a superconducting gap may open when a chiral 2DES is in proximity to a BCS superconductor substrate or electrode~\cite{Morpurgo-1,Morpurgo-2};
the $K$ and $K'$ valleys may couple to each other and become gapped in pair annihilation by Kekul\'{e} pattern of bond distortions~\cite{Hou-07,Luo-09}.
Nevertheless, our focus will be the chiral symmetry breaking and spontaneous gap (mass) generation~\cite{Zhang-11-SQH} that is driven by electron-electron interactions~\cite{Min-08-MF,Zhang-10-RG,Levitov-10-MF}.

\section{Spontaneous Symmetry Breaking}
Because of the large density-of-states and the $N\pi$ Berry phases
near low-energy band touching points, chiral 2DES with $N>1$ at zero external fields are strongly susceptible to broken symmetry states~\cite{Zhang-11-SQH,Sun-09,Vafek1,Levitov-10-SQH,Zhang-12-LAF}.
It is of interesting to determine whether the layer pseudospin orientations in model~(\ref{H0}) will be driven out-of-plane or acquire an in-plane distortion in the presence of electron-electron interactions~\cite{Min-08-MF,Guinea,Jung-11-MF,Trushin-11,MacDonald-12-MF}.
This amounts to asking for each spin-valley flavor whether the chiral symmetry or the rotational symmetry will be spontaneously broken.
In this section we review perturbative renormalization group (PRG) analysis~\cite{Zhang-10-RG,Sun-09,Vafek1,Guinea,Levitov-10-RG,Falko1,Vafek2,Falko2,fRG,MC,Vafek4,Vafek3,Zhang-12-RG,Kharitonov-12,Varma-13},
in which lattice effects are completely ignored and the long-range of the Coulomb interaction is not treated explicitly. The continuum approach is strongly motivated by the low-density of strongly correlated electrons.
The use of short-range interactions is crudely justified~\cite{Zhang-12-RG} by appealing to screening considerations,
and arguing that the momentum-independent interaction parameters represent an average over the relevant portion of momentum space.
Note that the lengthscale, related to the higher energy cutoff $\sim\gamma_1=400$ meV, is more than $10$ times of the graphene lattice constant $a$.
We therefore emphasize~\cite{Zhang-12-RG} that unlike the Hubbard model where the short-range interactions are the on-site repulsions, the short-range interactions approximation below correctly takes into account the long-range character of Coulomb interactions implicitly.
In our view models in which interactions are cut off at atomic length scales~\cite{Vafek2,Vafek3}, although technically interesting,
are unlikely to be relevant to few-layer graphene.

Interacting bilayer graphene behaves in many ways as if it is one dimensional electron system with linear dispersions,
because it has Fermi points instead of Fermi lines and because its particle-hole energies have a quadratic dispersion which compensates for the difference between 1D and 2D phase spaces~\cite{Zhang-10-RG}.
PRG analysis of bilayer graphene is inspiring when it is compared with the case for one-dimensional electron gas (spinless Luttinger liquid)~\cite{Zhang-10-RG,Sun-09,Levitov-10-RG,Zhang-12-RG}.
The main merit of the PRG is that it treats all virtual processes on an equal footing~\cite{Shankar-94}.
The spontaneous symmetry breaking in bilayer graphene can be most easily understood by considering only one spin-valley flavor~\cite{Zhang-10-RG,Sun-09,Zhang-12-RG}.
In such a spinless and valleyless model, Pauli exclusion principle allows only one antisymmetrized interaction parameter, i.e., the interlayer interaction $V_D$.
A PRG analysis determines how the bare interaction $V_D$ at the $\gamma_1$ scale is renormalized to $\Gamma_D$ at energy $\sim \gamma_1e^{-2\ell}$ by
integrating out the high-energy fermion degrees of freedom. A standard calculation yields the following PRG flow equation at the one-loop level
\begin{eqnarray}
\frac{d\,\Gamma_D}{d\,\ell}=\nu_0\,\Gamma_D^2\,,
\end{eqnarray}
where $\nu_0=\gamma_1/(4\pi\hbar^2v_0^2)$ is the constant density-of-states per flavor in bilayer graphene.
This flow equation implies a repulsive interaction instability toward a broken symmetry state~\cite{Zhang-10-RG,Sun-09,Vafek1}.

More insight into the likely nature of the broken symmetry state which occurs in bilayer graphene can be obtained by using the PRG calculations to estimate the long-wavelength static limit of the layer pseudospin susceptibilities~\cite{Zhang-12-RG}: $\chi_{\alpha\beta}$ with $\alpha,\beta=x,y,z$. The conservation of particle in the two layers at long wavelength implies $\chi_{\alpha\beta}\sim\delta_{{\alpha\beta}}$.
Divergences in $\chi_{zz}$ signal ordered states in which the charge density is transferred between layers for each flavor,
breaking the chiral symmetry and producing an energy gap for charged excitations~\cite{Zhang-11-SQH,Min-08-MF,Zhang-10-RG,Levitov-10-MF}.
Divergences in $\chi_{xx}$ or $\chi_{yy}$, on the other hand imply broken symmetry states in which the phase relationship between layers is altered and rotational symmetry within the layers is broken~\cite{Vafek1,Falko1,Geim-11}.
The broken rotational symmetry state is usually referred to as the gapless nematic state.
Provided that the system has a single continuous phase transition, the ordered state character can in principle be determined by identifying the pseudospin susceptibility which diverges first as temperature is lowered.

In the noninteracting case, the band state layer pseudospin structure yields susceptibilities~\cite{Zhang-12-RG}
\begin{eqnarray}
\chi_{zz}^0=2\chi_{xx}^0=2\chi_{yy}^0= 2\nu_0\ell\,.
\end{eqnarray}
These pseudospin susceptibilities capture the contribution of zero-momentum vertical
interband quantum fluctuations involving states with energies measured from the $K$ and $K'$ points between $\gamma_1$ and $\gamma_1e^{-2\ell}$.
The factor of two difference between $\chi_{zz}$ and $\chi_{xx}$ demonstrates that the band state is more susceptible to a gap opening perturbation than to a nematic perturbation. This property can be understood
in terms of the layer pseudospin orientations of the band states~\cite{MacDonald-12-MF}; in
particular a $\hat{z}$ pseudospin effective field is perpendicular to the valence-band pseudospin for
all momentum orientations, so that $\hat{z}$ direction fields always yield a strong response.
On the other hand $\hat{x}$-$\hat{y}$ plane pseudospin fields are not in general perpendicular to the
valence-band pseudospin and consequently produce a weaker response averaged over momentum
orientations.  The larger response to a $\hat{z}$ pseudospin effective field is related to the
well-known property~\cite{McCann-06} of bilayer graphene that a potential difference between the top and bottom layers lead to an energy gap at the $K$ and $K'$ points.

When the interactions flow to strong values, the susceptibilities are dominated by their interaction contributions.
Near the instability the renormalized interaction leads to
\begin{eqnarray}
\chi_{zz} = 4 \chi_{xx}\, =  4 \chi_{yy}\,,
\end{eqnarray}
suggesting that a gapped state breaking the chiral symmetry is most likely~\cite{Zhang-12-RG}.
Recent experiments~\cite{Bao-12,Yacoby-10-1,Yacoby-10-2,Schonenberger-12,Lau-12,Wees-12,Schonenberger-13,Lau-14,Others} also appear to rule out the competing family of gapless nematic states~\cite{Vafek1,Falko1,Geim-11}.
Hereafter we will neglect the non physical gapless nematic state breaking the rotational symmetry.
We point out that a more rigorous four-flavor model leads to results that are qualitatively similar,
although the PRG has a tendency to overestimate the instability toward the gapless nematic state~\cite{Zhang-12-RG}.
It is the layer pseudospin orientations of band states that frustrate the in-plane pseudospin distortion and favor the out-of-plane pseudospin alignment~\cite{Min-08-MF,MacDonald-12-MF}.

Similar spontaneously broken symmetry physics can also be generalized to thicker ABC-stacked $N$-layer graphene~\cite{Zhang-11-SQH,Zhang-12-LAF,Zhang-12-RG,Jung-13,Vafek-13,Jia-13}.
For larger $N$, the low energy bands are increasingly flat and the pseudospin chirality (Berry phase) is
larger, at least when weak remote hopping processes are neglected, leading to larger opening for many-body interaction effects.
In the simplified chiral model~(\ref{H0}), the density-of-states $\nu(E)\sim E^{(2-N)/N}$ diverges as $E$ approaches zero for $N>2$ whereas it remains finite for $N=2$ and zero for $N=1$. This difference corresponds to the fact that the short-range interactions at tree level are irrelevant for $N=1$, marginal for $N=2$, and relevant for $N>2$.
This indicates even stronger chiral symmetry breaking in $N>2$ films~\cite{Zhang-12-RG}.
In practice, however, the increased strength of trigonal warping would eventually dominate the decreased magnitude of interlayer exchange beyond a critical $N$.
In recent experiments interaction-driven spontaneous gaps $\sim 2$~meV and $\sim 50$~meV have been observed in suspended bilayers~\cite{Bao-12,Schonenberger-12,Lau-12,Wees-12,Schonenberger-13,Lau-14,Others}, trilayers~\cite{Lau2011,Lau2014},
and tetralayers~\cite{Mo2015} respectively,
whereas there has been no evidence for a charge gap in monolayers~\cite{mono-gap-1,mono-gap-2}.
Similar to the bilayer case, the competing ordered states in $N>2$ films are anticipated to be gapped states~\cite{Zhang-11-SQH,Zhang-12-LAF,Zhang-12-RG,Jung-13,Vafek-13,Jia-13}, which break the chiral symmetry.
Note that in these systems that have already been accessible in experiment, the many-body spontaneous gaps are larger than the single-particle trigonal warping energies.

\section{Berry Curvatures, Orbital Moments, and Hall Conductivities}
Now we discuss the electronic properties of $N$-layer chiral 2DES systems using the
ordered state quasiparticle Hamiltonians~\cite{Zhang-11-SQH} suggested by the above PRG analysis
\begin{eqnarray}
H_N^{\rm Int}=H_N+m\sigma_z\,,
\end{eqnarray}
where the mass term $m\sigma_z$, whose sign is valley- and spin-dependent, breaks the chiral symmetry and produces an energy gap.
we note that the Berry curvatures are non-zero only when chiral symmetry is broken~\cite{Zhang-14-Chiral,Zhang-13-Mirror}.
Indeed, the momentum-space monopoles, i.e., the original band-contact points, become gapped and the Berry curvature spreads out near the two valley centers. Since $m\ll\gamma_1$ the momentum dependence of $m$ can be neglected. The Berry curvature~\cite{Xiao-10} of the $N$-layer chiral 2DES is~\cite{Zhang-11-SQH}
\begin{eqnarray}\label{eq:berry}
\Omega^{(\alpha)}_{\hat z}({\bm p},\tau_{z},s_{z})=
-\alpha\frac{\tau_{z}}{2}\frac{m}{h_{t}^3}\bigg(\frac{\partial h_{\parallel}}{\partial p}\bigg)^2\,,
\end{eqnarray}
where the symbol $\alpha=+(-)$ denotes the conduction (valence) band, and the transverse and total pseudospin fields are $h_{\parallel}=(v_0
p)^N/\gamma_1^{N-1}$ and $h_{t}=\sqrt{m^2+h_{\parallel}^2}$. The orbital magnetic moment carried by a Bloch electron
is $M^{(\alpha)}_{\hat{z}}=e\hbar\epsilon^{(\alpha)}\Omega^{(\alpha)}_{\hat z}$ for a two-band model with particle-hole symmetry~\cite{Xiao-10}.
For the chiral 2DES the orbital moment is~\cite{Zhang-11-SQH}
\begin{eqnarray}\label{eq:berry}
M^{(\alpha)}_{\hat z}({\bm p},\tau_{z},s_{z})=\bigg[-\tau_{z}\frac{m}{h_{t}^2}\bigg(\frac{\partial
h_{\parallel}}{\partial p}\bigg)^2 m_{e}\bigg]\mu_{B}\,,
\end{eqnarray}
where $m_{\rm e}$ is the electron mass and $\mu_{\rm B}$ is the Bohr magneton $e\hbar/2m_{\rm e}$. Like the Berry curvature the orbital
magnetization changes sign when the valley label changes {\em and} when the sign of the mass term (the sense of layer polarization) changes,
{i.e.} both are proportional to $\tau_{\rm z} \rm{sgn}(m)$. The orbital magnetization is however independent of the band index
$\alpha$.

In the presence of an in-plane electric field, a quasi-electron acquires an anomalous transverse velocity proportional to the Berry curvature, giving
rise to an intrinsic Hall conductivity. Based on Eq.~(\ref{eq:berry}), the intrinsic Hall conductivity contribution from a given valley and spin is given by~\cite{Zhang-11-SQH}
\begin{eqnarray}\label{eq:hall}
\sigma_{\rm H}^{(\alpha)}(\tau_{z},s_{z})=\frac{\tau_{z}}{2}\frac{Ne^2}{h}\bigg(\frac{m}{h_{t}
\left( p_F \right)}-\frac{m}{|m|}\delta_{\alpha,+}\bigg)\,,
\end{eqnarray}
where $h_t (p_F)$ is the total pseudospin field at the Fermi wavevector. Provided that the Fermi level lies in the mass gap, each spin and valley
contributes $Ne^2/2h$ to the Hall conductivity, with the sign given by $\tau_{z} \rm{sgn}(m)$.
As we will review in the next section, different broken chiral symmetry states can be classified by their charge, spin, and valley Hall conductivities that are determined by the signs of $m$ for different valleys and spins~\cite{Zhang-11-SQH,Levitov-10-SQH}.
For this reason, the broken chiral symmetry states were often referred to as spontaneous quantum Hall states.

\section{Topological Classification}
\begin{figure}[t!]
\centering{ \scalebox{0.36} {\includegraphics*{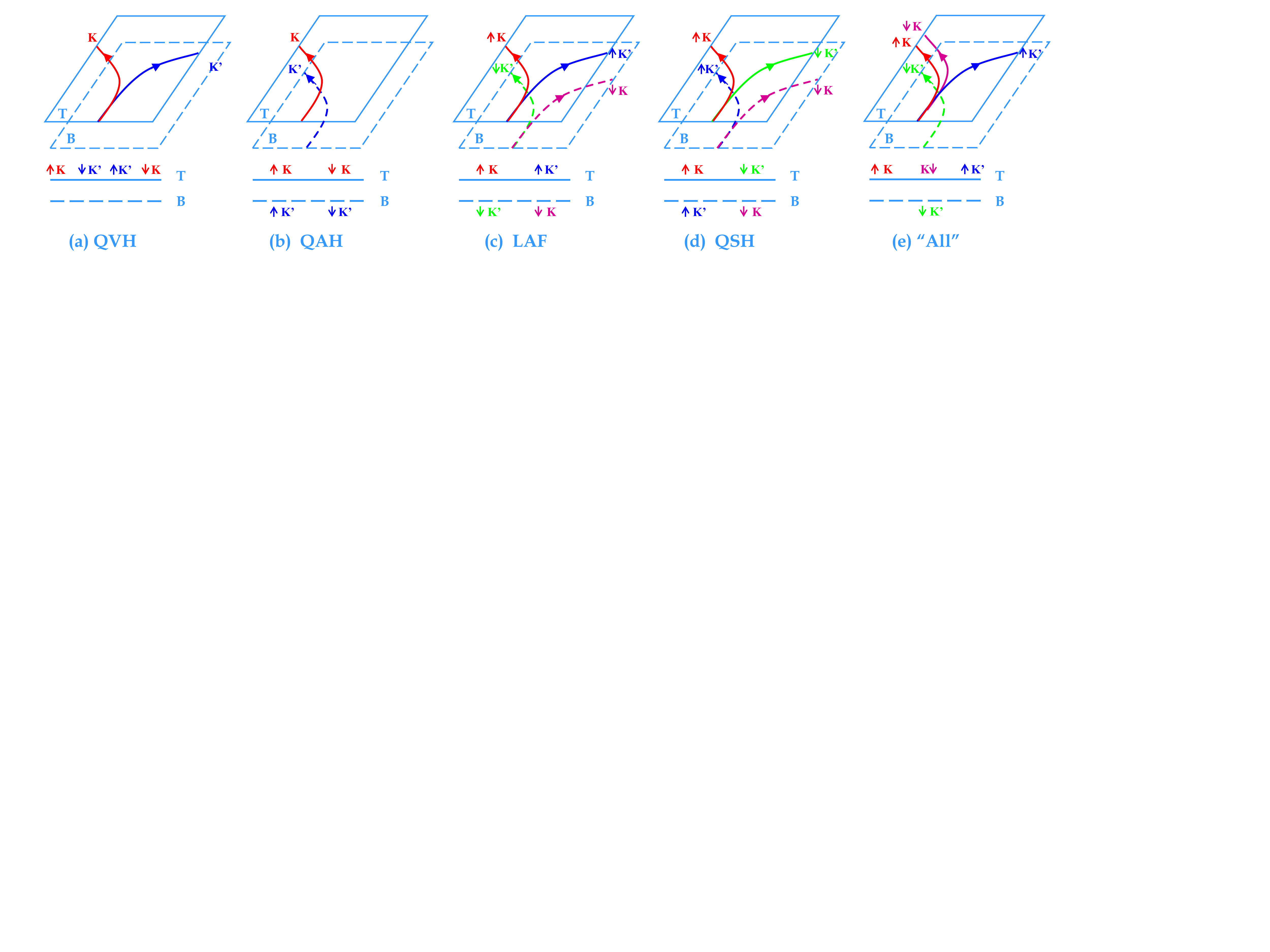}}}\caption{\label{5states} {
For cases (a-e) the lower panel describes the sense of layer polarization for each spin-valley combination,
whereas the upper panel schematically indicates the corresponding Hall conductivity contributions.
(a) A quantum valley Hall insulator with a net layer polarization and a spontaneous mass term $m\sigma_z$;
(b) an quantum anomalous Hall insulator with a valley-dependent spontaneous mass term $m\tau_z\sigma_z$;
(c) a layer-antiferromagnetic insulator with a spin-dependent spontaneous mass term $m\,s_z\sigma_z$;
(d) a quantum spin Hall insulator with a valley- and spin-dependent spontaneous mass term $m\tau_zs_z\sigma_z$;
(e) an exotic spontaneous quantum Hall state with a valley and spin dependent spontaneous mass term
$m(\frac{1+\tau_{z}}{2}+\frac{1-\tau_{z}}{2}s_{z})\sigma_{z}$. Figures adapted from refs.~\onlinecite{Zhang-11-SQH,Zhang-11-DW}.}}
\end{figure}

When spin is ignored only two different types of states can be distinguished: ones in which the $K$ and $K'$ valleys are layer polarized in the opposite sense producing a quantum anomalous Hall (QAH) state with broken time-reversal symmetry ($\Theta$) and orbital magnetization, and ones in which the two valleys have the same sense of layer polarization producing a quantum valley Hall (QVH) state with broken inversion symmetry ($\mathcal{P}$) and zero total Hall conductivity~\cite{Zhang-11-SQH,Levitov-10-SQH}. When spin is included, there are three additional states~\cite{Zhang-11-SQH}. The five distinct states in the spinful case can be obtained by each spin choosing to be a QVH state or a QAH state. Below we will explain how these states are distinguished by their charge, spin, and valley Hall conductivities, by their orbital magnetizations, by their broken symmetries. These results are sketched in Fig.~\ref{5states} and summarized in Table~\ref{table:one}.

\subsection{Quantum valley Hall state}
In Fig.~\ref{5states}(a), we consider the case in which each spin-valley is polarized in the same sense. The total Hall conductivity is then zero for both spins, with the $K$ and $K'$ valleys making Hall conductivity and magnetization contributions of opposite sign.
This state preserves time-reversal symmetry but breaks inversion symmetry.
Besides the layer polarization, the inversion symmetry breaking yields a valley Hall effect, analogous to the quantum spin Hall effect~\cite{Kane-05}.
The spontaneous mass term in this case can be modeled by $m\sigma_z$ which is spin-valley independent.
Compared to the other four states, the QVH state has more layer polarization and hence stronger Hartree (electrostatic) energy cost~\cite{Min-08-MF,Jung-11-MF}.
The QVH states can be favored experimentally by applying a perpendicular electric field,
which breaks the inversion symmetry and produces the layer polarization.

\subsection{Quantum anomalous Hall state}
Fig.~\ref{5states}(b) describes the case in which the two valleys are polarized in the opposite sense, toward the top and bottom layers respectively.
The spontaneous mass term can thus be modeled by $m\tau_z\sigma_z$ which is spin independent but valley dependent.
This implies Hall conductivity and orbital moment (or orbital magnetization) contributions of the same sign for each spin and valley. This inversion symmetric state, breaks time-reversal symmetry and the valley Ising ($\mathcal{Z}_2$) symmetry, but its spin density is surprisingly everywhere zero. The total Hall conductivity has the quantized value $2Ne^2/h$ and we refer to this state as the QAH state~\cite{Haldane,Qiao}.
In addition to its QAH effect, this state has a substantial orbital magnetization~\cite{Zhang-11-SQH}.
The QAH state, without total layer polarization, is probably most simply identified experimentally by observing a $\nu=2N$ quantum Hall effect which persists to zero magnetic field.
Such a state can be favored experimentally by applying a perpendicular magnetic (orbital) field, which couples the orbital moments and hence lowers the energy~\cite{Zhang-11-SQH}.

\begin{table}[b!]
\centering
\caption{Summary of spin-valley layer polarizations (T or B) and corresponding charge, spin, and valley Hall conductivities
($e^2/h$ units) and insulator types for the three distinct states (b-d) with no overall layer polarization, for a state in which every
spin-valley is polarized toward the top layer (a), and for a state with partial layer polarization (e).
The last column indicates the symmetries that are broken by each state.
Table adapted from refs.~\onlinecite{Zhang-11-SQH,Zhang-12-LAF}.}
\newcommand\T{\rule{0pt}{3.1ex}}
\newcommand\B{\rule[-1.7ex]{0pt}{0pt}}
\begin{footnotesize}
\centering
\begin{tabular}{c | c c c c || c | c | c | c | c | c | c}
\hline\hline Figure & $K\uparrow$ & $K\downarrow$ & $K'\uparrow$ & $K'\downarrow$
& $\sigma^{\rm (SH)}$ & $\sigma^{\rm (VH)}$ & $\sigma^{\rm (CH)}$ & $\sigma^{\rm (SVH)}$ &
Insulator & Mass Operator & Broken Symmetries \T\\[3pt]
\hline\T
\ref{5states}(a)  & T & T & T &  T  &  $0$ & $2N$ & $0$ & $0$ & QVH & $m\sigma_z$ & $\mathcal{P}$ \\[3pt]
\ref{5states}(b)  & T & T & B & B  &  $0$  &  $0$ & $2N$ & $0$ & QAH & $m\tau_z\sigma_z$ & $\Theta$,\,$\mathcal{Z}_2$\\[3pt]
\ref{5states}(c)  & T & B & T & B  &  $0$  &  $0$ & $0$ & $2N$ & LAF & $m s_z\sigma_z$ & $\Theta$,\,$\mathcal{P}$,\,$SU(2)$\\[3pt]
\ref{5states}(d)  & T & B & B & T  &  $2N$ & $0$  &  $0$ & $0$ & QSH & $m\tau_z s_z\sigma_z$ & $\mathcal{Z}_2$,\,$SU(2)$\\[3pt]
\ref{5states}(e)  & T & T & T &  B  & $N$ & $N$ & $N$ & $N$ & All & $m(\frac{1+\tau_{z}}{2}+\frac{1-\tau_{z}}{2}s_{z})\sigma_{z}$ & $\Theta$,\,$\mathcal{P}$,\,$\mathcal{Z}_2$,\,$SU(2)$\\[3pt]
\hline\hline
\end{tabular}
\end{footnotesize}
\label{table:one}
\end{table}

\subsection{Layer antiferromagnetic state}
For the state depicted in Fig.~\ref{5states}(c), there is also no net charge transfer between the layers,
as the two spins are layer polarized in the opposite sense.
Such a state can be viewed as the two spins having QVH effects of opposite sign and the two layers having spin polarizations of opposite sign.
Thus, its charge, spin, and valley Hall conductivities are all zero, and only the spin-valley Hall conductivity is nontrivial.
The spontaneous mass term takes the form of $m s_z\sigma_z$, which is valley independent but spin dependent.
Evidently, this state breaks time-reversal symmetry, inversion symmetry, and the spin $SU(2)$ symmetry.
In sharp contrast to the ordinary AF state or spin-density-wave state that occurs in the lattice scale,
the spontaneous state considered here, driven by the long-range Coulomb interactions, only has a sizable interlayer spin density macroscopically.
For instance, the spin density is negligible for each low-energy site. For these reasons, this state was dubbed as a LAF state by the author originally~\cite{Zhang-11-SQH,Zhang-12-LAF}. As we will discuss in a short while, the LAF state is the ground state at zero external fields, favored by the intrinsic intervalley exchange interaction~\cite{Jung-11-MF} which is rather small.

\subsection{Quantum spin Hall state}
Fig.~\ref{5states}(d) depicts a state in which there is no layer polarization, and neither the top nor the bottom layer has spin or valley polarization. Quite interestingly, if we consider only one layer, there are both spin Hall and valley Hall effects; however, the orientations of the Hall currents in the top and the bottom layers are the same (opposite) for the spin (valley) Hall effects.
While breaking the spin $SU(2)$ symmetry and the valley $\mathcal{Z}_2$ symmetry, this state does not break inversion and time-reversal symmetries, and instead has QAH effects of opposite signs in the two spin subspaces.
The spontaneous mass, may be viewed as an effective spin-orbit coupling (SOC), takes the form of $m s_z\tau_z\sigma_z$,
leading to a quantized spin Hall conductivity.

The spontaneous QSH effect is in several respects~\cite{Zhang-11-SQH} different from that discussed in the well known papers~\cite{Kane-05}, which foreshadowed the identification
of $\mathcal{Z}_2$ topological insulators. (i) The QSH effect here is driven by broken symmetries produced by electron-electron interactions, rather than by SOC, which we neglect. The effective SOC due to electron-electron interactions can be $10^4$ times larger than the intrinsic one. (ii) Unlike the previous interaction-induced QSH phase, which appears only at finite interaction strengths~\cite{Raghu-08,Wen-10}, here the instability to the QSH phase is present even for weak interactions. The instability occurs only for $N>1$ systems rather than in the monolayer system. (iii) The spontaneous QSH phases are topologically characterized by the spin (or mirror) Chern number with integer values, rather than by a $\mathcal{Z}_2$ label. Of course, the $N$-odd layers are $\mathcal{Z}_2$ topological insulators.

\subsection{``All'' state}
Among the five possible distinct states, the most interesting is the one in which one spin-valley flavor polarizes in the opposite sense of the other three, as sketched in Fig.~\ref{5states}(e). This state has a QVH effect for the spin up flavor and a QAH effect for the spin down flavor.
This state can also be understood as a valley-filtered QAH state~\cite{Pan}.
Remarkably enough, charge, valley, and spin Hall effects coexist in this state~\cite{Zhang-11-SQH}.
For this reason, we dubbed such a state ``All'' state.
The interaction induced mass term for the state sketched in Fig.~\ref{5states}(e) may be modeled by $m(1+\tau_{z}+s_z-\tau_{z}s_{z})\sigma_{z}/2$.
Note that there are eight different ``All'' states,
and their mass terms may be modeled by switching the signs of $\tau_z$, $s_z$, and/or $\sigma_z$ in $m(1+\tau_{z}+s_z-\tau_{z}s_{z})\sigma_{z}/2$.
The ``All'' state should be favored in the presence of an interlayer electric field and a finite magnetic field~\cite{Zhang-11-SQH,Lau-14},
because its energy is lowered by the orbital and spin coupling to the magnetic field and by the compensation of the Hartree
energy cost of its layer polarization by the electric field.

\subsection{Lattice effects and ground states}
The above five states are rather close in energy, e.g., their energy differences are only a few percent of their condensation energies~\cite{Min-08-MF}.
When lattice effects are taken into account, {\it intervalley exchange} weakly favors states in which the sense of layer polarization is identical for both valleys of either spin~\cite{Jung-11-MF}. This effect, which is not accounted above, favors the LAF state.
More intriguingly, as we have discussed and will show below, distinct states or their adiabatic variations can be realized by tuning the external electric or/and magnetic fields.

\section{Experimental Evidences in Bilayer Graphene}
Recently, suspended bilayer graphene devices can be fabricated to exhibit ultra-high mobilities $\geq 10^5$~cm$^2$/(Vs).
This made possible a series of delicate measurements in quantum Hall regime and in transport spectroscopy near charge neutrality~\cite{Yacoby-10-1,Yacoby-10-2,Schonenberger-12,Lau-12,Wees-12,Schonenberger-13,Lau-14,Others}.
As we will review below, there have been strong experimental evidences for four of the five gapped states.
(Note that the unobserved QSH state is similar to the $\nu=0$ state at large perpendicular magnetic field when the bilayer graphene is put on substrates. In this sense, one may argue the observation~\cite{Maher-13} of QSH state, with a quantized spin Chern number, although time-reversal symmetry is explicitly broken.)

\subsection{Local compressibility}
\begin{figure}[t!]
\centering{ \scalebox{0.52} {\includegraphics*{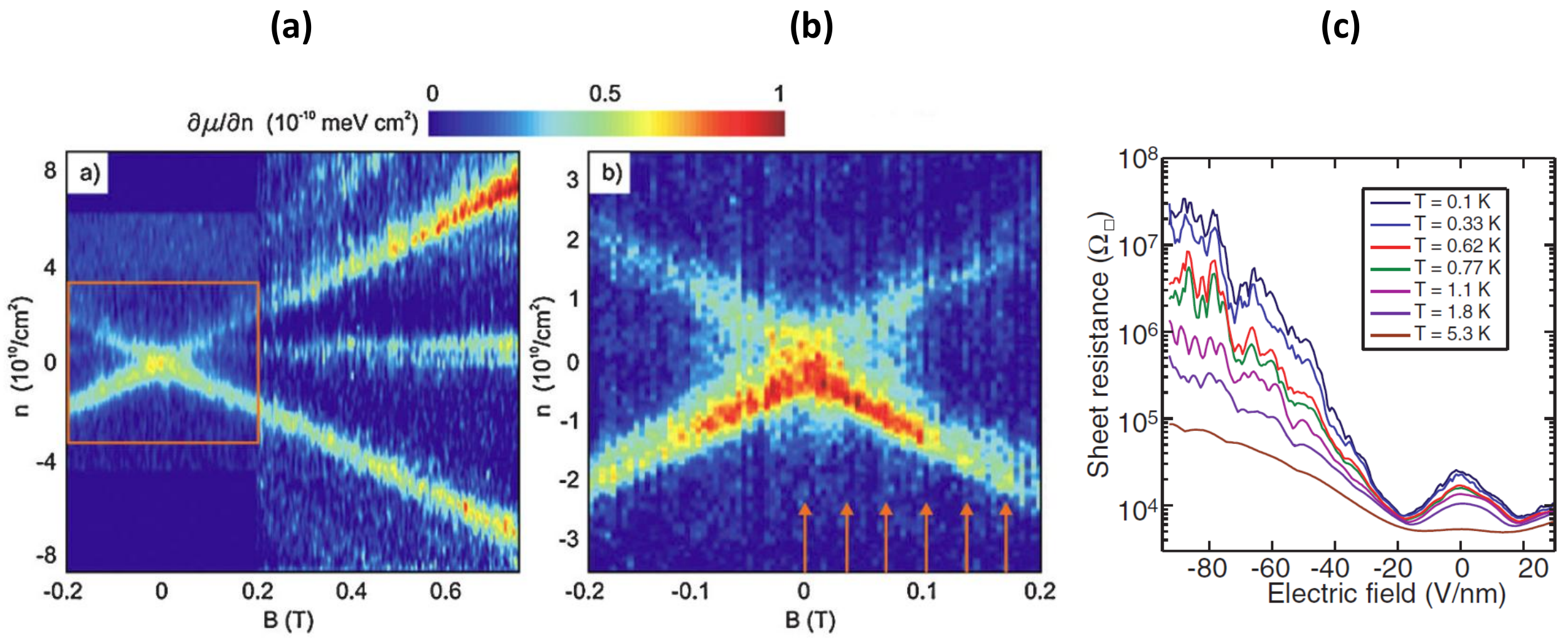}}}\caption{\label{exp1} {
(a) The local inverse compressibility measurement in single gated suspended bilayer graphene as a function of the magnetic field at zero electric field. (b) The zoom in image of (a) near the zero magnetic field. (c) The sheet resistance of dual gated suspended bilayer graphene as a function of the interlayer electric field and the temperature at zero magnetic field. Figures adapted from refs.~\onlinecite{Yacoby-10-1, Yacoby-10-2}.}}
\end{figure}

The first experimental evidence~\cite{Yacoby-10-1} of spontaneous quantum Hall states was from the local inverse compressibility measurement in
single gated suspended bilayer graphene
near charge neutrality. As shown in Fig.~\ref{exp1}(a) and (b), the $\nu=0$ and $\nu=4$ integer quantum Hall states persist down to nearly zero magnetics fields. This unprecedented observation is suggestive of the existence of broken symmetry states with Hall conductivities $0$ and $4e^2/h$. Consistent with the earlier analysis that the QAH states can be favored by a small perpendicular magnetic field that couples to the orbital moment. Similar $\nu=4$ states have also been observed at nearly zero field by several other groups.

Fig.~\ref{exp1}(c)~\cite{Yacoby-10-2} further shows transport measurements in dual gated suspended bilayer graphene. At zero magnetic field, the resistivity is larger than that of the non interacting case. More interestingly, as the interlayer electric field with either polarity increases, the resistivity decreases first and then increases with no drop. This phenomenon is consistent with the quantum phase transition between a QVH state and a gapped ground state without total layer polarization, although the gap nature is not conclusive here.

\subsection{Transport spectroscopy}
\begin{figure}[t!]
\centering{ \scalebox{0.55} {\includegraphics*{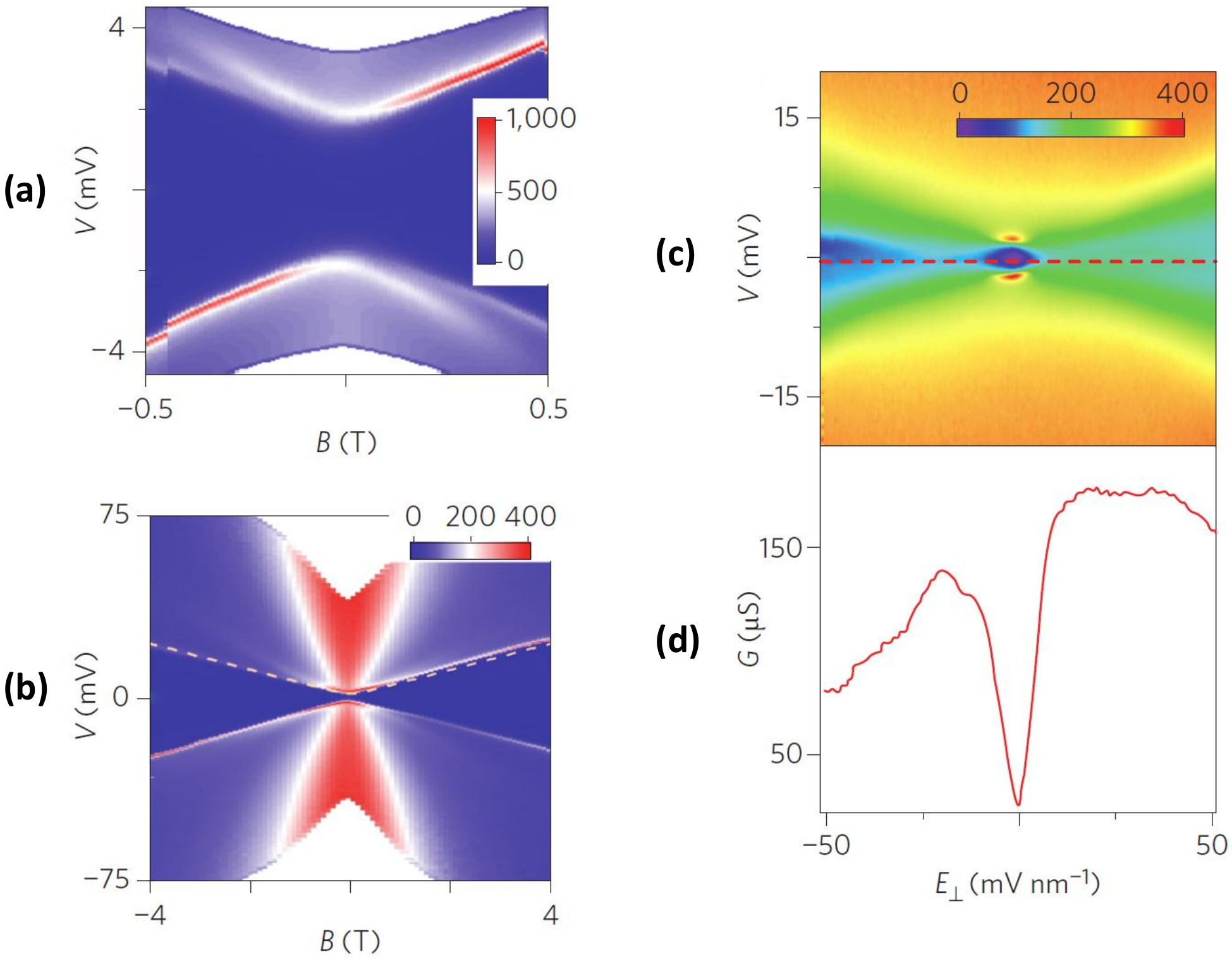}}}\caption{\label{exp2} {
(a) and (b) show the two-terminal conductance $G$ as a function of the magnetic field $B$ and the source-drain bias $V$ for charge neutral bilayer graphene at the interlayer electric field $E_\bot=0$.
(c) and (d) show the two-terminal conductance $G$ as a function of the interlayer electric field $E_\bot$ for charge neutral bilayer graphene at the magnetic field $B=0$. Figures adapted from refs.~\onlinecite{Lau-12}.}}
\end{figure}

More conclusive evidences for the gap nature and the ground state class have been observed based on reproducible~\cite{Bao-12} two-terminal conductance measurements in dual gated bilayer graphene at charge neutrality~\cite{Lau-12}. As shown in Fig.~\ref{exp2}(a) and (b), the two-terminal conductance at zero magnetic field and zero electric field is nearly zero for a wide window of the source-drain bias. Clearly, the conductance is much smaller than $\sim e^2/h$ of the non interacting case. This strongly suggests a charge gap $\sim 2$~meV in bilayer graphene at zero field and zero density.

At zero magnetic field in Fig.~\ref{exp2}(c) and (d), as the interlayer electric field with either polarity increases from zero, the conductance increases quickly first and then drop gradually. This implies a quantum phase transition from a state without layer polarization to a QVH state that is favored in the presence of an interlayer potential difference.

Combining the evidences in Fig.~\ref{exp2}(a)-(d), one can conclude that the ground state of bilayer graphene has zero charge Hall conductivity, no protected edge states (since the total conductance is almost zero), and no layer polarization. The asymmetry features at the positive and negative magnetic fields further indicate time-reversal symmetry breaking. All these experimental observations are consistent with the prediction that the ground state is likely a LAF state. When the magnetic field is turned on, no matter it is in-plane or out-of-plane, the LAF state becomes a canted AF state~\cite{Zhang-12-LAF,Kharitonov-12}.

\subsection{Evidence for the ``All'' state}
\begin{figure}[t!]
\centering{ \scalebox{0.4} {\includegraphics*{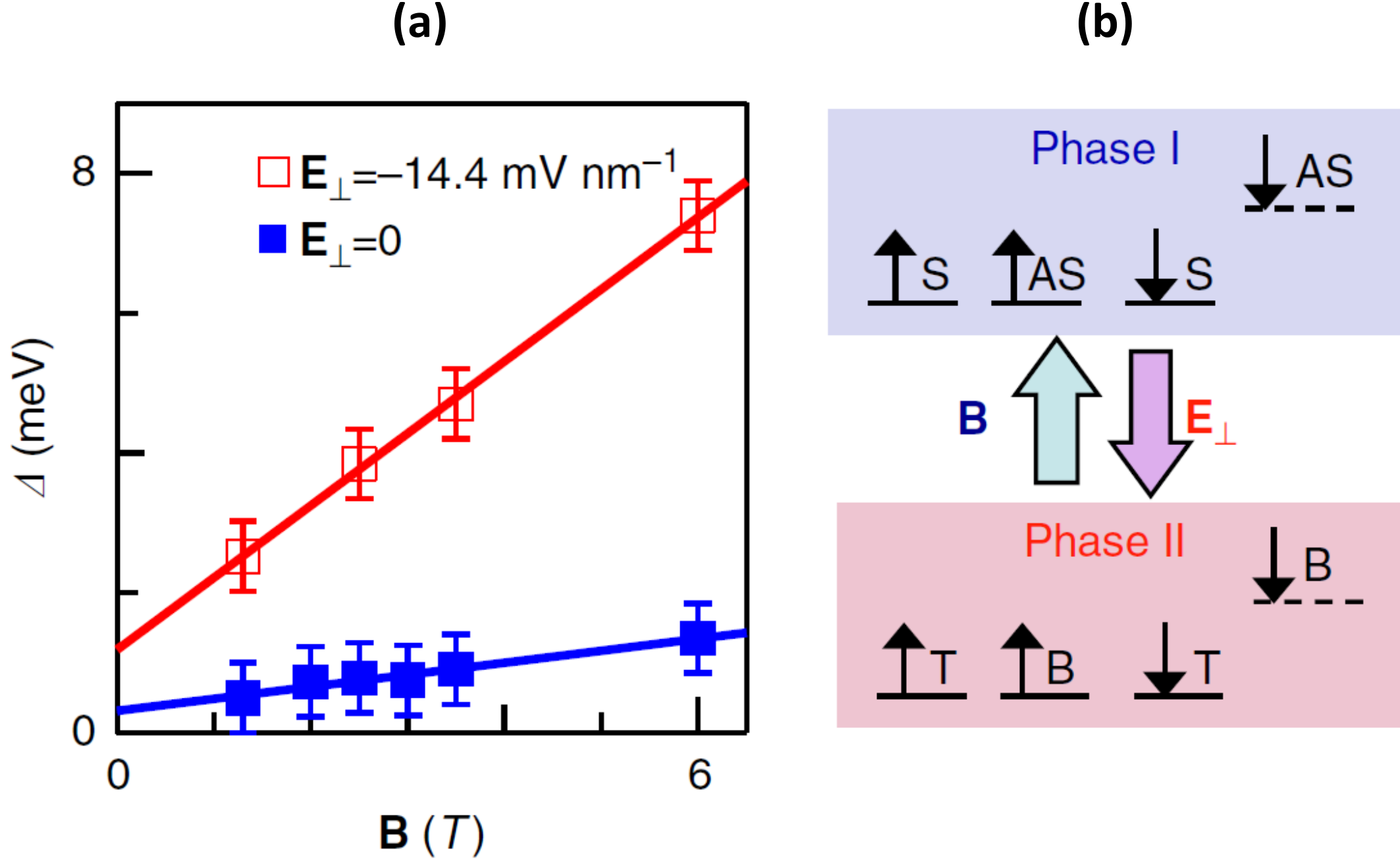}}}\caption{\label{exp4} {
(a) Measured bilayer graphene Landau level gap $\Delta(B)$ at $\nu=2$ and $E_\bot=0$ and $E_\bot=-14.4$~mV/nm, respectively.
Error bars are estimated from bias resolution and uncertainties in the valley width measurement.
(b) Schematics of transitions between phases I and II at $\nu=2$.
(T/B, top/bottom layer; S/AS, symmetric/antisymmetric state; $|S\rangle=|T\rangle+|B\rangle$, $|AS\rangle=|T\rangle-|B\rangle$.)
Figures adapted from refs.~\onlinecite{Lau-14}.
}}
\end{figure}

Most recently, the ``All'' state (see Fig~\ref{5states}(e) and Table~\ref{table:one}) has been observed experimentally in the quantum Hall regime in the presence of an interlayer electric field~\cite{Lau-14}. In Fig.~\ref{exp4}, two $\nu=2$ states are observed in suspended bilayer graphene.

Phase I is only fully resolved at zero electric field and large magnetic fields, with a vanishing Landau level gap at zero magnetic field.
The fact that a large B is required to stabilize this phase has been anticipated in the quantum Hall ferromagnetism theory of bilayer graphene~\cite{Barlas-12}.
As phase I is only observed in the absence of electric field, it is evidently not layer polarized but a symmetric linear combination of the top and bottom layers, or the $K$ and $K'$ valleys equivalently, and thus a valley-Kekul\'{e} state~\cite{Lau-14}.

Remarkably in Fig.~\ref{exp4}(a), in contrast to phase I, phase II is observed at anomalously small magnetic fields and large electric field,
with a Landau level gap that extrapolates to a finite zero-magnetic-field intercept. Its appearance at much smaller magnetic field than phase I is
reminiscent of the ``All'' state at zero magnetic field. Indeed phase II is only metastable at zero external fields, most likely because it loses the ordering competition to the LAF state at zero electric field and to the QVH state at finite electric fields. As observed in Fig.~\ref{exp4}~\cite{Lau-14} and predicted earlier~\cite{Zhang-11-SQH}, however, phase II can be preferred in the presence of both finite electric and finite magnetic fields, since states with different total
Hall conductivity are most stable at different carrier densities; moreover, the energy of this phase is lowered by the orbital and
spin coupling to the magnetic field and by the compensation of the Hartree energy cost of its layer polarization by the electric field. We emphasize that this state is partially polarized in spin, valley, and layer, as shown in Fig~\ref{5states}(e) and Table~\ref{table:one}.

\subsection{Comparison to mean-field theory}
When broken chiral symmetry physics occurs within each spin and valley of a chiral 2DES, as suggested by the PRG and susceptibility analysis,
the spontaneous gap $2m$ at zero carrier density can be obtained by solving the self-consistent mean-field gap equation~\cite{Zhang-12-LAF}
\begin{eqnarray}
\label{eqn:Ngap}
m=\frac{V}{2A}\sum_{\bm p}\frac{m}{\sqrt{\epsilon_{\bm p}^2+m^2}}\,,
\end{eqnarray}
$V$ is the effective interaction strength between
electrons near the $K$ or $K'$ valleys, and $A$ is a sample area.

For the monolayer case with $N=1$, the instability occurs for $V\nu_1^*>2$,
where $\nu_1^*=W/(2\pi\hbar^2v_0^2)$ is the Dirac-model density-of-states at the ultraviolet energy cutoff scale $W\sim \hbar v_0/a$.
Thus, the symmetry breaking would only occur for large interaction strengths.
We can conclude from this observation that the presence or absence of a broken symmetry state in monolayer graphene depends on atomic length scale physics beyond that captured by the chiral model~\ref{H0} with $N=1$. It has been established experimentally that the gaps, if present, cannot be larger than $0.1$~meV in monolayer graphene~\cite{mono-gap-1,mono-gap-2}.

In sharp contrast to the $N=1$ case, the gap equation~(\ref{eqn:Ngap}) always has a solution for any $N>1$ case, no matter how small $V$ is.
This is consistent with the weak repulsive interaction instability that is concluded by the PRG and susceptibilities analysis~\cite{Zhang-12-RG}.
We will focus on the bilayer case below.
In suspended ultra-clean bilayer graphene at zero carrier density,
where $m$ is small compared to $\gamma_1$, the solution of the gap equation is given accurately by~\cite{Zhang-12-LAF}
\begin{eqnarray}
m=2\gamma_1 e^{-{2}/{\nu_0 V}}\,.
\label{eqn:mass}
\end{eqnarray}
The spontaneous gap $2m=2$ meV observed in experiment at zero magnetic field corresponds to
a dimensionless interaction strength $\nu_0 V=0.2992$, close to the value expected to be appropriate for screened Coulomb interactions~\cite{mono-gap-1,mono-gap-2}.

\begin{figure}[t!]
\centering{ \scalebox{0.50} {\includegraphics*{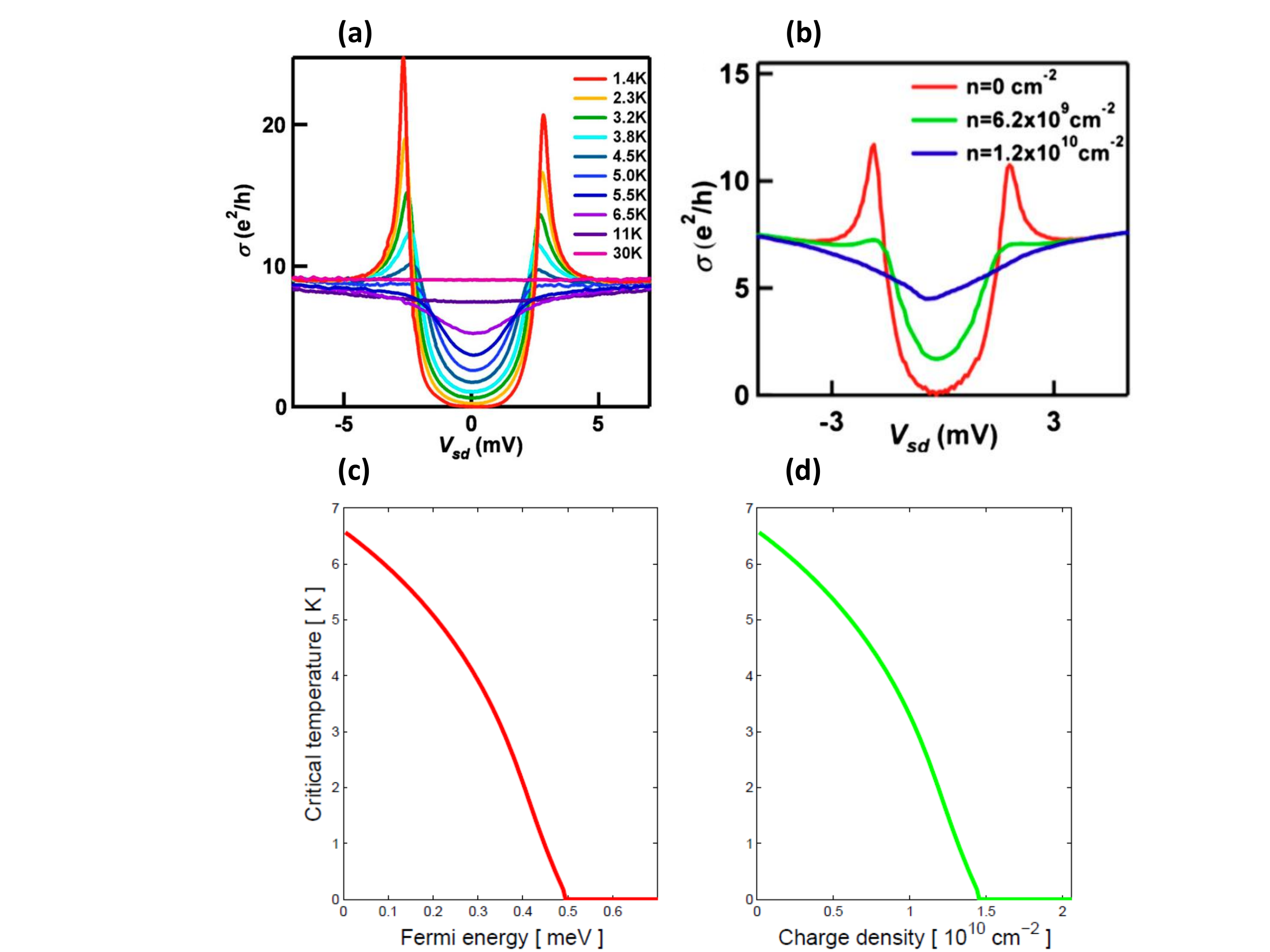}}}\caption{\label{exp3} {
(a) The temperature-dependent two-terminal conductance $\sigma$ as a function of the source-drain bias $V_{sd}$ in bilayer graphene at the lowest carrier density $n$.
(b) The carrier-density-dependent two-terminal conductance $\sigma$ as a function of the source-drain bias $V_{sd}$ in bilayer graphene at the lowest temperature $T$.
(c) and (d) are the mean-field estimates of the critical temperatures $T_c$ as a function of the fermi energy or the carrier density.
Figures adapted from refs.~\onlinecite{Bao-12,Supp-Bao}.}}
\end{figure}

When the bilayer graphene carrier density is nonzero, it is possible to show that the temperature $T=0$ mass is reduced to~\cite{Supp-Bao}
\begin{eqnarray}
m'=\sqrt{(m-\epsilon_{F})^2-\epsilon_{F}^2}\,,
\end{eqnarray}
because of Pauli blocking effects in the gap equation. Mean-field theory therefore predicts that the broken symmetry state disappears once Fermi energy becomes larger than $m/2=0.5$~meV, which corresponds to a carrier density larger than
\begin{eqnarray}
n_{c}=1.47\times 10^{10}~\mbox{cm}^{-2}\,.
\end{eqnarray}
As compared in Fig.~\ref{exp3}, both the size of the gap at zero carrier density and the critical carrier density
which destroys the gap are therefore roughly consistent with simple mean-field-theory estimates~\cite{Bao-12,Supp-Bao}.
The smallness of the density fluctuation that destroys the gapped state is consistent with its appearance only in suspended samples.

The measured $T_{c}$ in Fig.~\ref{exp3}(a) (as well as in Fig.~\ref{exp1}(c)) at zero charge density and $n_{c}$ at zero temperature are in good agreement with the mean-field theory results in Fig.~\ref{exp3}(c).
The critical temperature at finite carrier density can be further determined by solving~\cite{Supp-Bao}
\begin{eqnarray}
\label{eqn:Tc}
\frac{2}{\nu_0 V}=\int_{0}^{\gamma_1}\frac{f_{c}(-\epsilon-\mu)-f_{c}(\epsilon-\mu)}{{\epsilon}} d\epsilon\,,
\end{eqnarray}
Where $f_{c}(\epsilon)=(1+e^{\epsilon/k_{B}T_{c}})^{-1}$ is the Fermi function at critical temperature $T_{c}$. The solution for $T_{c}$ as a function of chemical potential and carrier density is plotted in Fig.~\ref{exp3}(d).

\section{Discussion \& Outlook}
In conclusion, the mean-field-theory estimates and the experimental observations in bilayer graphene are in reasonable agreement with each other.
The spontaneous chiral symmetry breaking in chiral 2DES has provided a new paradigm for many-body physics, i.e., the spontaneous quantum Hall effects near zero magnetic field~\cite{Zhang-11-SQH}. Below we will briefly discuss a few predictions and phenomena that are relevant to the physics reviewed in this article.

\subsection{Edge states}
The physical significance of spontaneous charge, valley, and spin anomalous Hall effects also exhibits in their edge state spectra~\cite{Zhang-11-SQH}.
Evidently, the QAH (QSH) states should have protected chiral (helical) edge states corresponding to their bulk topological invariants.
Similarly, the ``All'' states should also have protected chiral edge states, though the number is reduced by half.
Analogous to the helical edge states of QSH states, the QVH as well as the LAF state have counterpropagating edge states at different valleys.

Importantly, the QVH edge states are not strictly topological and can be gapped out by a sufficiently strong, large-momentum scattering which couples the two valleys, even if the underlying symmetries are still preserved~\cite{Zhang-13-PNAS}. It is therefore crucial that the valley index remains a good quantum
number, for which short-range disorder, interlayer stacking, and electric fields must be smooth on the lattice scale. Under this condition, backscattering is prohibited and the QVH edge states become protected and thus attractive.
Graphene edges often have sufficiently strong atomic scale defects that liberate intervalley scattering, and the edge states are thus destroyed. However, one can create a smooth domain wall that separates two different QVH states and observe robust edge states along the domain wall, since the smoothness of domain wall would prohibit intervalley scattering.
For example, QVH-like edge states were predicted~\cite{Zhang-13-PNAS,Vaezi} to exist along a domain wall that separates an AB and a BA stacked domains in bilayer graphene under an uniform interlayer electric field. Recently, such QVH domain wall modes have been observed~\cite{Ju2015} in a transport measurement.

\subsection{Domain walls}
Bilayer graphene is susceptible to a family of unusual broken chiral symmetry states that are rather close in energy.
Thus, domain walls, in which the sense of layer polarization of at least one flavor is reversed, are expected to be present in disordered bilayer graphene samples.
A recent calculation~\cite{Li-14} shows that the metal-insulator transition temperature in bilayer graphene is reduced from mean-field estimates by thermal excitation of domain walls.
The same calculation also demonstrates that the domain walls have interesting microscopic structure related to the topological character of the
ordered states~\cite{Li-14}.
It turns out there are $16$ distinct types of domain walls in the spinful case and $3$ in the spinless case, each of which has topological domain wall modes~\cite{Zhang-11-DW,Li-14}. These unusual domain wall modes are anticipated to exhibit novel Luttinger liquid behaviors~\cite{Zhang-11-DW,Affleck, LL}.

\subsection{Distinguishing distinct states}
While the three gapped states without layer polarization respond similarly to an interlayer electric field, which can induce first order transitions at which the total layer polarization jumps, they have distinct responses to Zeeman coupling to their spin~\cite{Zhang-12-LAF}.
When an in-plane magnetic field is included, the QAH state quasiparticles simply spin-split, leaving the ground state unchanged but the charge gap
reduced to $2m-g\mu_B B$. For a $2$~meV spontaneous gap at the zero field, a field of $17$~T drives the gap to zero.
For the LAF state, the in-plane magnetic field induces a noncollinear spin state in which the components of the spin-density perpendicular to the field are opposite in opposite layers, while those along the field direction grow smoothly with field strength and are identical.
Most importantly, the gap does not change as the Zeeman field increases.
As indicated by the mean-field mass terms above, a QSH state contains two copies of LAF states with opposite signs for different valleys.
In a recent experiment~\cite{Schonenberger-13}, the spontaneous gaps of bilayer graphene have been found robust to the Zeeman fields, consistent with that earlier confirmation that the ground state is the LAF state.
However, this experiment only employed in-plane magnetic fields up to $3$~T, which is not sufficiently large enough to influence the observed ground states with gaps $\sim 2$~meV. A future experiment with a larger in-plane magnetic field could provide more conclusive evidences.

The QAH state and the QVH state are found to have interesting and distinct optical absorption of circularly polarized lights with a frequency close to the spontaneous gap. Both valleys of the QAH state only absorb lights with one helicity, whereas different valleys of the QVH state absorb lights with the opposite helicities~\cite{Trushin-11,Trushin2,Yao-08}. One can further analyze the optical absorption for other three broken symmetry states, since they can be viewed as the two spin flavors choosing to be different QVH and/or QAH states. Since the two spin flavors break time-reversal symmetry in the same sense, the QAH state is the only one that can be distinguished by Kerr optical spectroscopy~\cite{Levitov-11,Gusynin-12}.

\subsection{Quantum Hall effects}
On one hand, each spontaneous quantum Hall state at zero magnetic field is adiabatically connected to a quantum Hall ferromagnet with the same charge Hall conductivity~\cite{Zhang-11-SQH,Levitov-10-SQH,Zhang-12-LAF}. That said, there is a smooth evolution between the zero-field and high-field broken symmetry states in suspended chiral $N$-layer 2DES with $N>1$.
On the other hand, even for chiral 2DES put on substrates, where the chiral symmetry does not occur at zero magnetic field, the quantum Hall ferromagnetism~\cite{Zhang-12-QHF,Barlas-12} still occur at sufficiently high magnetic fields. Moreover, an interlayer electric field can also drive quantum phase transitions between different quantum Hall ferromagnets. What is unprecedented is the fact that integer quantum Hall effects can occur in AB bilayers and ABC trilayers way below $0.1$~T. What is more exotic would be the fractional quantum Hall effects near zero magnetic field, which is still elusive.

\subsection{Kondo insulators, flat-band insulators, and van der Waals heterostructures}
Finally, we briefly comment on three closely related phenomena.
Topological Kondo insulators~\cite{Dzero-10} have been discovered in $\rm SmB_6$ recently. The underlying mechanism is rather similar to the spontaneous QSH states of chiral 2DES with $N>1$. Near the Fermi energy the interaction induces an energy gap between the two already inverted bands, a nearly localized-flat $f$-band and a $d$-derived dispersive conduction band, leading to a transition from a Kondo lattice metal to a small-gap topological Kondo insulator.
In a similar sense, the flat-band Chern insulators driven by interactions are similar to the spontaneous QAH states of chiral 2DES with $N>1$.
An intriguing task is to identify {\it realistic} systems that host topological phases with fractionalized excitations~\cite{Z4}, beyond the celebrated fractional quantum Hall states.
Recently, two dimensional van der Waals heterostructures constitute a new class of artificial materials formed by stacking atomically thin planar crystals. It is likely that exotic broken symmetry physics may occur when one layer is a chiral 2DES and the other layers are materials with small dielectric constants.

\begin{acknowledgments}
I would like to thank my main collaborators on this topic, A. H. MacDonald and C. N. Lau.
I am also indebted to inspiring discussions with many other collaborators on few-layer graphene: W. Bao, G. A. Fiete, J. Jung, C. L. Kane, X. Li, E. J. Mele, H. Min, Q. Niu, H. Pan, M. Polini, Z. Qiao, D. Tilahun, J. Velasco Jr., B. J. Wieder, S. A. Yang, J. Zhu, and K. Zou.
This work was made possible by support from UT Dallas research enhancement funds.
\end{acknowledgments}


\begin{thebibliography}{999}
\bibitem{Weyl}
 A Weyl point is two-fold degeneracy near which the effective Hamiltonian can be written in the form a two-component Weyl Hamiltonian with either left or right chirality. In contrast, a Dirac point is a four-fold degeneracy near which the effective Hamiltonian can be written in the form of massless four-component Dirac Hamiltonian. A Dirac point can be viewed as two Weyl points with opposite chirality sit on top of each other. Therefore, a $K$ or $K'$ point in a graphene system is better called as a spin-degenerate Weyl point, since the two spin species have exactly the same chirality.
\bibitem{Chiral-Note}
 This was originally named as a layer-inversion symmetry for a single flavor, because the breaking of this symmetry yields charge transfer between the top and bottom layers for the considered flavor. Yet, when all the four flavors are taken into account, the spatial inversion symmetry may or may not be broken, as shown in Table~\ref{table:one}.
\end{thebibliography}
\end{document}